\documentclass[12pt]{article}
\usepackage{epsfig,mfigure}
\newcommand{ \mpx }{\langle p_{x} \rangle}

\begin{document}

\title{Anisotropic Flow at STAR} 

\author{
R.J.M.~Snellings,
A.~M.~Poskanzer
and
S.~A.~Voloshin} 

\date{March 2, 1999}
\maketitle

\begin{abstract}
We first review previous work on anisotropic flow at the AGS and
SPS. Then the physics related to flow is discussed as well as the
interaction of flow with other non-flow measurements. From 40k RQMD
and 100k HIJING events predictions for anisotropic flow at RHIC are
presented. Using the STAR detector acceptance, estimates for the
resolution obtainable with STAR are shown. We conclude that it should
be possible to obtain good measurements for elliptic flow with either
the STAR main TPC or forward TPCs. Anisotropic flow should be easily
one of the first results from STAR.
\end{abstract}
\section{Introduction}

\vspace{-13.cm}
\hfill
{\Large STAR Note SN388}
\vspace{13.cm}

The study of collective flow in nuclear collisions at high energies
has been attracting increasing attention from
experimentalists~\cite{H-G,olli98}. This is partly because recent
progress has been made in the development of new techniques suitable
for flow studies at high
energies~\cite{olli92,lvzh,ollimeth,olli98,posk98,meth}. Instead of
studying $\mpx$, in these new methods a Fourier expansion of the
azimuthal distribution of particles is used in which the first
harmonic coefficient, $v_1$, quantifies the directed flow and the
second harmonic coefficient, $v_2$, quantifies the elliptic flow. In
some cases $A1$ and $A2$ were reported, which in modern terminology,
are twice the square of the sub-event resolution. Using these new
techniques anisotropic flow has now been observed for heavy symmetric
systems at both the AGS and SPS.

At the AGS the E877 Collaboration pioneered the use of the Fourier
expansion method to measure $v_1$ and $v_2$. They studied these
quantities (as well as $v_4$) from a calorimeter as a function of
centrality in different pseudorapidity windows~\cite{l877flow1}. Then
they studied nucleons as well as pions as a function of pseudorapidity
for different centralities~\cite{l877flow3}. Using their spectrometer
to identify particles while still obtaining the event plane from the
calorimeter, they measured $v_1$ and $v_2$ as a function of $p_t$ for
different rapidities and centralities~\cite{l877flow2}. They also
reported $\mpx$ as a function of rapidity~\cite{l877flow2}. In their
latest papers they extended this study to light
nuclei~\cite{volo98,l877flow4}. The E802 Collaboration studied $\mpx$
for light nuclei in the target rapidity region using a forward
hodoscope to determine the event plane~\cite{ahle98}.

At the SPS NA49 first observed elliptic flow in a calorimeter study
which reported $A2$ as a function of centrality~\cite{lna49}. WA98
reported $A1$ as a function of centrality for protons and $\pi^+$ in
the target rapidity region~\cite{lwa98,wa9898}. They also studied
$\mpx$ in the target rapidity region~\cite{wa9898}. NA45 used silicon
drift detectors to study $v_1$ and $v_2$ as a function of
pseudorapidity~\cite{lna45}. NA49 has presented a differential study
of $v_1$ and $v_2$ as a function of $p_t$ and $y$~\cite{posk98} and
has also started to study the centrality
dependence~\cite{posk_annu99}.

Also, the importance of flow for other measurements has just begun to
be studied. For two particle correlations relative to the event plane
the mathematical scheme has been worked
out~\cite{lvc,wied97,heisel2,heisel}. Some first results have been
given by WA98~\cite{lwa98}. Also, for non-identical particles the
correlation relative to the event plane has been
discussed~\cite{volo97}.

\section{Physics Motivation}

Anisotropic flow, in particular elliptic flow, in spite of the
relatively small absolute value of the effect, contains very rich
physics.  In general words, it is very sensitive to the equation of
state which governs the evolution of the system created in the nuclear
collision.  Being such, anisotropic flow provides important
information on the state of matter under the extreme conditions of the
nuclear collision.  The anticipated phase transition to QGP should
have a dramatic effect on elliptic flow due to the softening of the
equation of state.

First it was pointed out in the pioneering work of
Ollitrault\cite{olli92}, who suggested elliptic anisotropy as a
possible signature of transverse collective flow. Within the
hydro-dynamical model Ollitrault analyzed the role of different
equations of state and phase transitions on the final anisotropy.
Hung and Shuryak~\cite{lshur} suggested scanning with beam energy in
order to look for the QCD phase transition. Using their idea of the
softest point in the equation of state combined with hydro-dynamical
calculations, Rischke~\cite{risc96} predicted a dramatic drop in the
elliptic flow signal at the corresponding beam energies (in the
original calculations this was at AGS energies).  Sorge has
shown\cite{lsorge} that the elliptic flow is very sensitive to the
pressure at maximum compression, which is the most interesting time in
the system evolution.  Recent studies~\cite{zhang99} within the parton
cascade model yield similar conclusions providing also the relation
between the strength of the elliptic flow and parton-parton cross
sections.  Recently, Sorge also tried~\cite{sorge98} to combine the
early system evolution in accordance to a QGP equation of state with a
later hadron cascade.  He looked at the centrality dependence of the
elliptic flow in order to detect QGP production.  Summarizing this
part, we would conclude that the effect of QGP should be seen in the
anisotropic flow dependence on the energy of the colliding nuclei, or
in the dependence on the centrality of the collision.  If the
situation would be such that a QGP is produced only in a small
fraction of the collisions than fluctuations in flow would be one of
the best observables for this effect.

The formation of DCC in nuclear collisions could also result in an
event anisotropy. It could be due to the anisotropic shape of the DCC
domains~\cite{wang98} or just to local fluctuations in the charged
multiplicity, which should result in ``orthogonal'' flow in charged
and neutral sectors~\cite{nayak}.

The very magnitude of anisotropic flow is sensitive to the degree of
equilibration in the system. Note that at present there is no
calculation based on the hydro-dynamical picture which accounts for
the experimentally observed values of the effect.  This could have its
origin in the obvious difficulties of hydrodynamic model calculations,
but it could also indicate a non-applicability of the picture to
nuclear collisions.  The cascade models such as RQMD describe the data
much better.  From this point of view the analysis of elliptic flow in
the collision-less and hydrodynamic limits performed in~\cite{heisel}
is very interesting.  The HBT interferometry performed relative to the
event plane~\cite{lvc,wied97,heisel2,heisel,volo_ann98_hbt} becomes
also extremely important at this point.  Does the system really expand
in the reaction plane as prescribed by hydrodynamics?  Simultaneous
measurements of the anisotropic flow and the two-particle, identical
as well as non identical~\cite{volo97}, correlations in principle
should answer this question.

We must also mention the importance of anisotropic flow measurements
to the vast variety of other measurements, which from first look have
nothing to do with anisotropic flow.  Let us consider high $p_t$
particle production. It could be that the production mechanism (hard
parton scattering) is very insensitive to the in-plane expansion, but
that the rescattering of high $p_t$ partons is different in the
different directions of particle emission due to the anisotropic
geometry of the collision zone.  This would lead to anisotropy in high
$p_t$ particle production and gives another opportunity to study how
it develops~\cite{snell_annu99,volo_ww99}. 

Another example is HBT measurements averaged over all orientations of
particle emission.  One would think that this does not require
reaction plane measurements, but this is not really true.  The mixed
pair distribution usually used in the correlation function calculation
can strongly depend on the relative orientation of the reaction plane
of the events used to create the mixed pair. Therefore one should have
this information even in the case where the dependence of the HBT
parameters on the reaction plane is not studied.

\section{Technical Requirements}

The study of azimuthal anisotropy of unidentified charged hadrons
needs the momenta of the particles but does not have any unusual
requirements for calibrations, momentum resolution, acceptance,
efficiency, two-track resolution, or two-track efficiency. However,
for future analyses it would be good to have particle identification.

\section{Directed and Elliptic Flow at RHIC}

The anisotropy in the azimuthal distribution of particles is often
characterized by $v_1$, $v_2$ and called directed and elliptic flow
respectively. This anisotropy, especially $v_2$, plays an important
role in high energy nuclear collisions and is expected to be even more
important at RHIC energies~\cite{lsorge}.  The azimuthal distribution
of particles is described by a Fourier expansion~\cite{lvzh}
\begin{equation}
E\frac{{\mathrm d}^3N}{{\mathrm d}^3p} =
\frac{1}{2\pi}
\frac{{\mathrm d}^2N}{p_t{\mathrm d}p_t{\mathrm d}y} \left(
1+\sum_{n=1}^{\infty} 
2 v_n \cos [n(\phi-\Psi_r)]\right),
\end{equation}
where $\Psi_r$ is the true reaction plane angle.  The reaction
plane is defined by the beam direction and the impact parameter vector
${\bf b}$.  In a given rapidity ($y$) and $p_t$ interval the
coefficients are determined by
\begin{equation}
v_{n} = \langle \cos [n(\phi-\Psi_r)] \rangle.
\end{equation}
Similarly this Fourier expansion can be done in coordinate space,
where for a given rapidity and $p_t$ interval the coefficients are
determined by
\begin{equation}
r_{n} = \langle \cos [n(\arctan (\frac{y}{x})-\Psi_r)] \rangle
\end{equation}
where $x,y$ are the particle space coordinates at freeze-out. Of
course, these equations only apply to simulations where one knows
$\Psi_r$.

Comparing the anisotropy coefficients in momentum space ($v_n$) with
the anisotropy coefficients in coordinate space ($r_n$) as a function
of $p_t$ helps us to understand the space-time evolution of
nucleus-nucleus collisions~\cite{lvc,nxu_vspace}.  To study this
space-time evolution at RHIC, $40\;000$ Au+Au collisions at $\sqrt
s$~=~200~$A$GeV have been analyzed using the RQMD v2.4 model.
\begin{figure}[t]
\centering
\mbox{
\mfigure[
{\it Anisotropy coefficients for nucleons and charged pions in RQMD
for collisions in the impact parameter range of 5 $\leq b \leq$ 10
fm.}  ] {\epsfig{figure=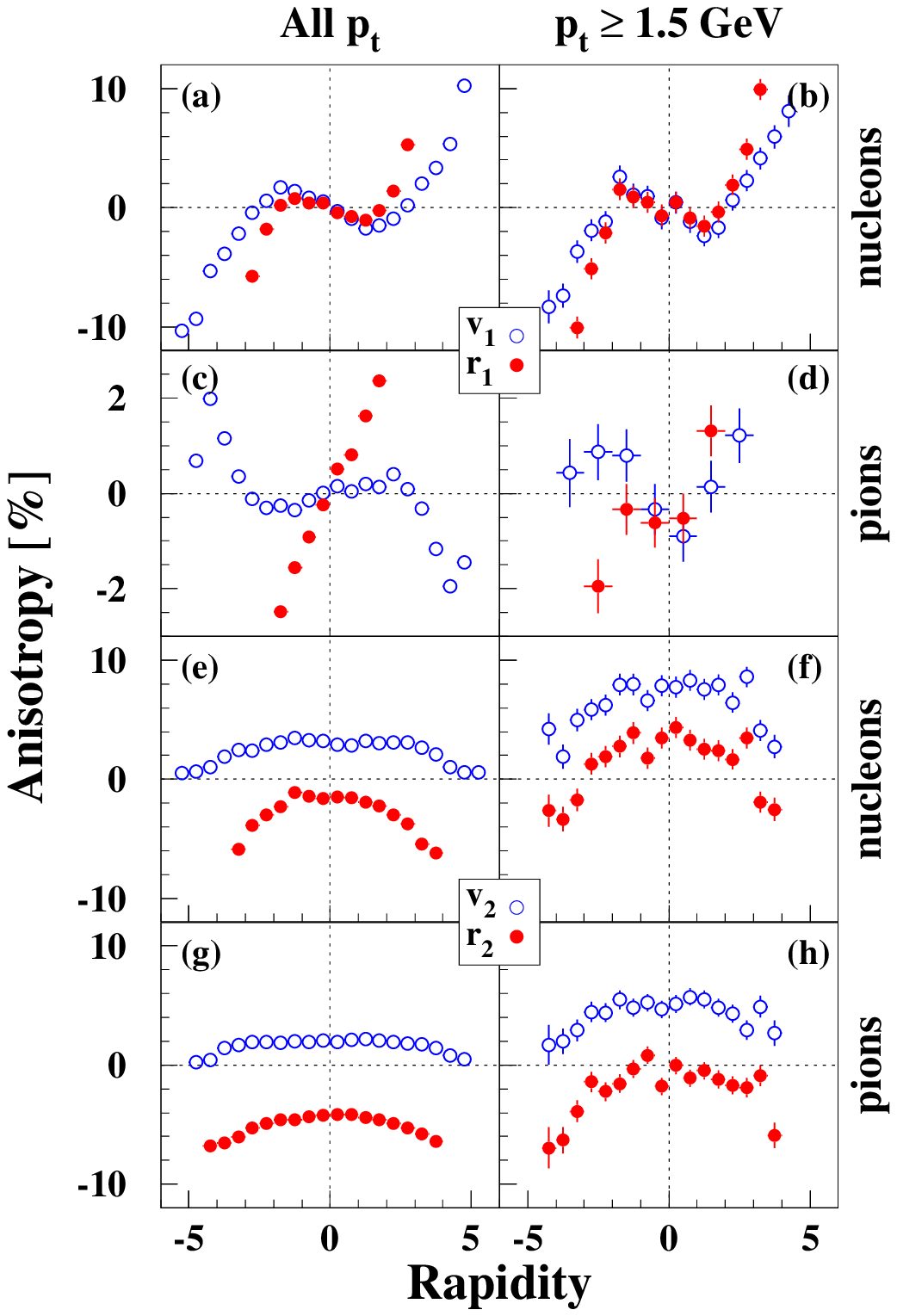,width=.49\textwidth}
\label{vxrx_rqmd}
}
\mfigure[
{\it Anisotropy coefficients for nucleons and charged pions in HIJING
for collisions in the impact parameter range of 5 $\leq b \leq$ 10
fm.}  ] {\epsfig{figure=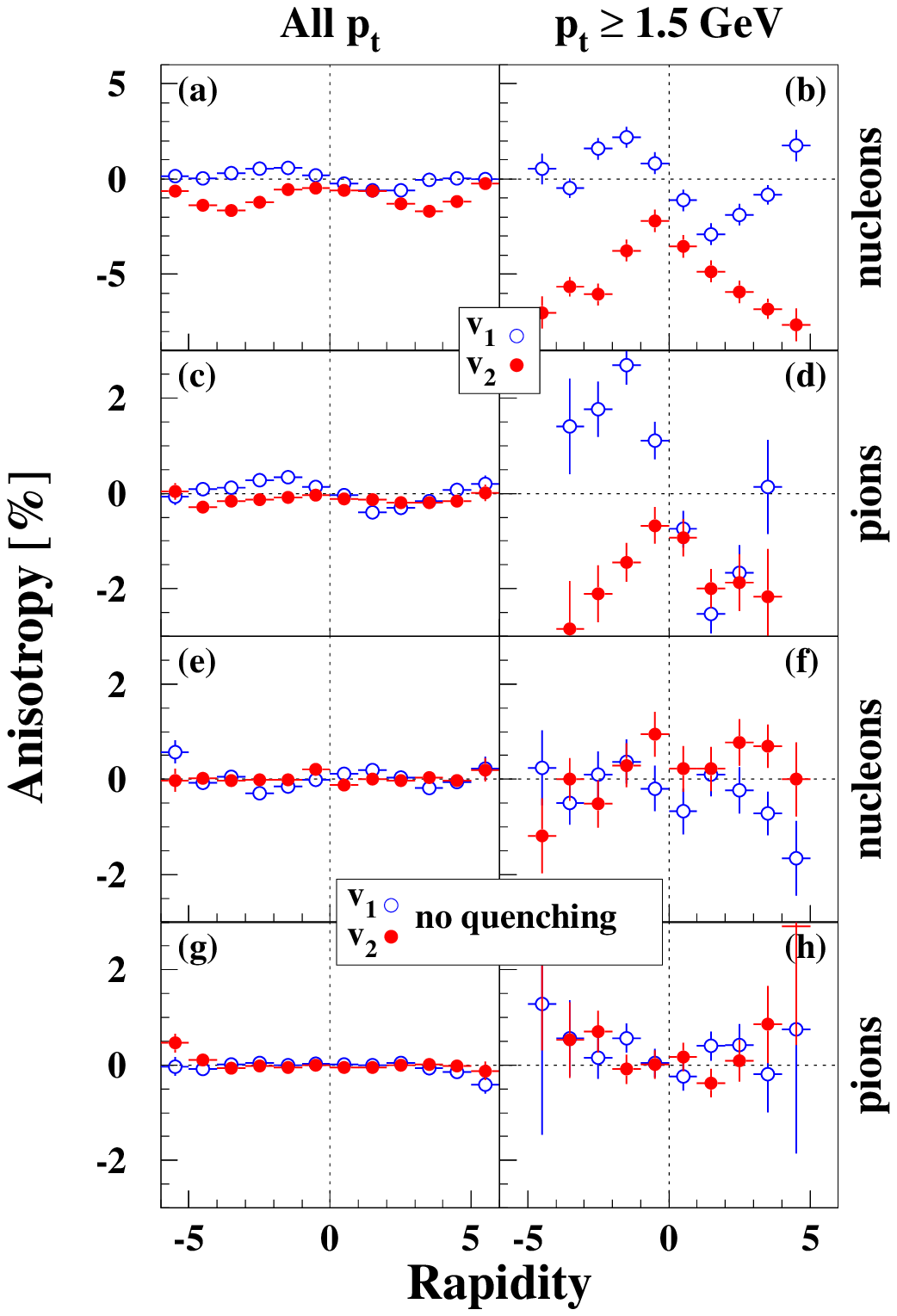,width=.49\textwidth}
\label{vx_all_hijing}
}
}
\end{figure}

Figs.~\ref{vxrx_rqmd}a-d show the first harmonic both in momentum and
coordinate space for nucleons and pions.  For nucleons at mid-rapidity
note the similarity in shape of $v_1$ versus $y$ and r$_1$ versus $y$.
Here (Fig. 1a) both the slopes of $v_1$ versus $y$ and r$_1$ versus
$y$ show a reversal of sign.  This finds an explanation in a picture
with strong (positive) space-momentum correlations, taking into
account the correlation between nucleon stopping and the original
position of the nucleons in the transverse plane~\cite{snell2_annu99}.
For pions, the rapidity dependence of $v_1$ is predominantly governed
by rescattering on comoving nucleons.  Figs.~\ref{vxrx_rqmd}e-h show
$v_2$ for nucleons and pions.  For both nucleons and pions $v_2$ is
positive and is larger for particles with $p_t \ge 1.5$ GeV. Particles
acquire a large $p_t$ when they are produced by a hard collision
(which should not produce an event anisotropy) or when they have a
large number of soft collisions (rescattering).  The latter would
explain the increase in $v_2$ and it explains why r$_2$ goes from
negative for nucleons integrated over all $p_t$ to positive for
nucleons with large $p_t$.

Collective flow and the coefficients $v_1$ and $v_2$ are usually
associated with soft processes. However, the coefficients describe the
event anisotropy and are not limited to only soft physics.  At RHIC
energies hard processes become important. They happen early in the
reaction and thus can be used to probe the early stage of the
evolution of a dense system.  During this time a quark-gluon plasma
(QGP) could exist.  Associated with hard processes are jets. However,
when the transverse energy of the jets becomes smaller it becomes
increasingly difficult to resolve them from the ``soft''
particles. These jets with $E_T <$ 5 GeV are usually refered to as
mini-jets.  At RHIC energies it has been estimated that 50\% of the
transverse energy is produced by mini-jets~\cite{jets}.

Medium induced radiative energy loss of high $p_t$ partons (jet
quenching) could be very different in a hadronic medium and a partonic
medium.  Recently it was shown that this energy loss per unit
distance, $dE/dx$, grows linearly with the total length of the
medium~\cite{baieretall}.  For non central collisions the hot and
dense overlap region has an almond shape. This implies different path
lengths and therefore different energy loss for particles moving in
the in-plane versus the out-plane direction.  To study this anisotropy
with respect to the reaction plane~\cite{snell_annu99}, $100\;000$
Au+Au collisions at $\sqrt s$~=~200~$A$GeV have been generated using
HIJING~\cite{hijing} v1.35.

Figs.~\ref{vx_all_hijing}a-d show $v_1$ and $v_2$ for nucleons and
charged pions. The coefficient $v_1$ shows a small negative slope
around mid-rapidity for both nucleons and pions and this becomes more
pronounced for particles with $p_t \ge 1.5$ GeV.  The coefficient
$v_2$ is slightly negative over the whole rapidity range for both
charged pions and nucleons. For particles with $p_t \ge 1.5$ GeV,
$v_2$ becomes more negative especially at forward and backward
rapidity. Figs.~\ref{vx_all_hijing}e-f show that without jet
quenching the anisotropy coefficients become zero. This indicates that
interactions among particles, either quenching or rescattering, are
important for producing the anisotropy.

\section{Event Plane Resolutions}

Within event generators the true reaction plane angle $\Psi_r$ is
known.  This is not the case experimentally and the reaction plane has
to be estimated from the data.  This is done using the anisotropy in
the azimuthal distribution of particles itself.  The estimated
reaction plane angle for the $n^{th}$ harmonic is called $\Psi_n$.
The magnitude of the anisotropy and the finite number of particles
available to determine this event plane leads to a finite resolution.
Therefore, the measured $v_n^{obs}$ coefficients with respect
to the event plane have to be corrected for this event plane
resolution
\begin{equation}
v_n = \frac{v_{n}^{obs}}{\langle \cos [n(\Psi_n-\Psi_r)] \rangle} \;.
\label{vn}
\end{equation}
However, eq.~\ref{vn} uses the true reaction plane which is not
known experimentally.  Following Ref.~\cite{meth}, if one constructs
the event plane from two random subevents one can relate the
resolution of the subevents to the full event plane resolution,
\begin{equation}
\langle \cos [n(\Psi_{n}-\Psi_r)] \rangle = 
C \times \sqrt{\langle \cos [n(\Psi_{n}^{a}-\Psi_{n}^{b})] \rangle }, 
\end{equation}
where $C$ is a correction~\cite{meth} for the difference in subevent
multiplicity compared to the full event and $\Psi_{n}^{a},
\Psi_{n}^{b}$ are the angles of the event planes determined in the
subevents.

\begin{figure}[t]
\centering
\mbox{
\mfigure[
{\it RQMD v2.4 prediction for $v_2$ using $\pi^+ +\pi^-$ within 
-1.5~$\leq y \leq$~1.5. The multiplicity and event plane resolution
are also shown.}
]
{
\epsfig{figure=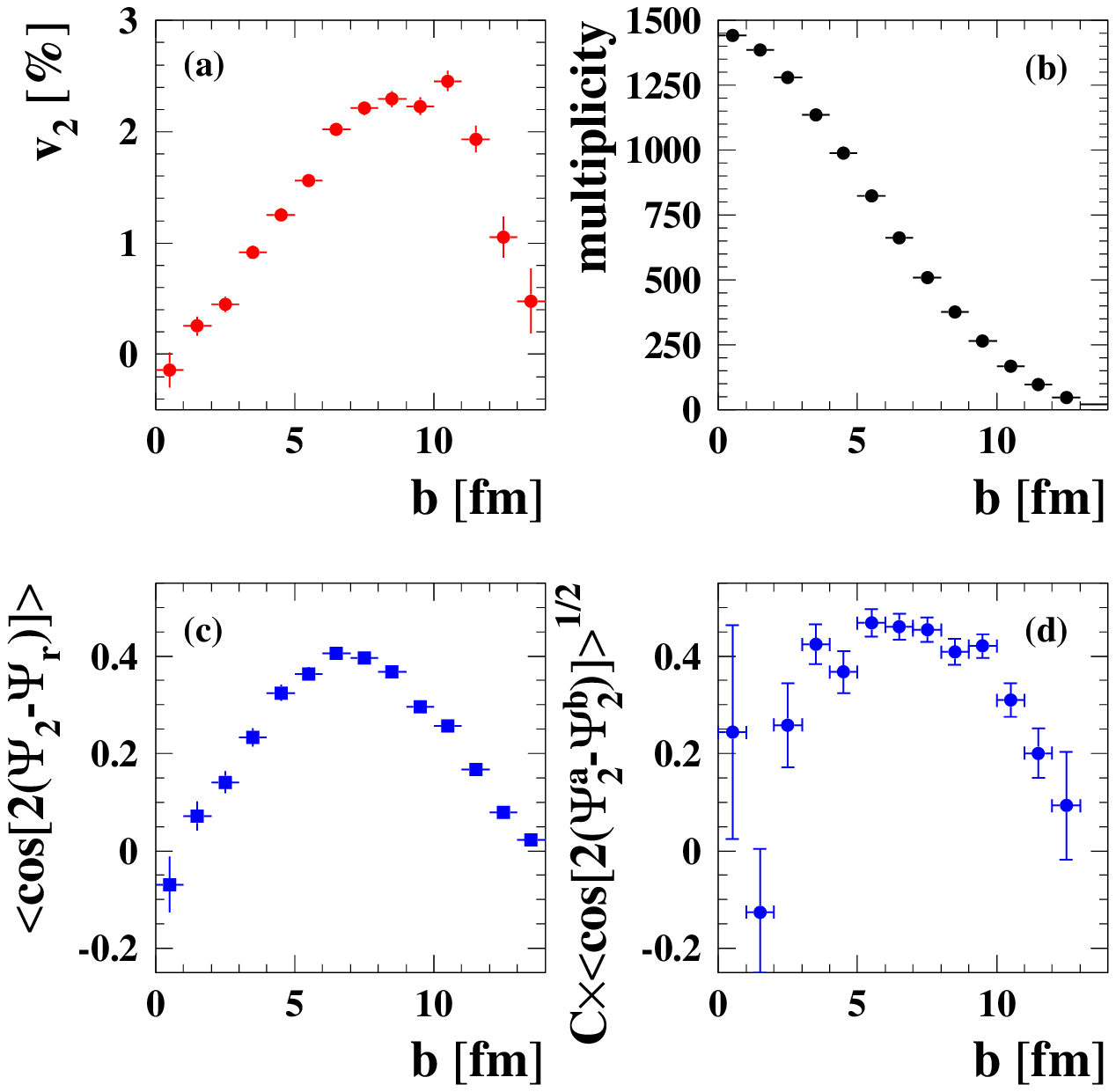,width=.68\textwidth}
\label{restpc}
}
}
\end{figure}

To calculate how well the event plane can be determined in STAR, we
considered the TPC (-1.5~$\leq y \leq$~1.5) and the FTPCs (2.5~$\leq
|y| \leq$~4.).  For this the RQMD v2.4 model predictions for Au+Au at
$\sqrt s$~=~200~$A$GeV have been used.  In Fig.~\ref{restpc}a, $v_2$
for charged pions integrated over the TPC rapidity region is shown
versus the impact parameter $b$.  Fig.~\ref{restpc}b shows the
corresponding multiplicity as a function of $b$.  These quantities
lead to a resolution for $v_2$, calculated using the true reaction
plane, as shown in Fig.~\ref{restpc}c.  The resolution for $v_2$ which
can be obtained in the STAR TPC using subevents is shown in
Fig.~\ref{restpc}d.  For $v_2$ charged pions and protons both
contribute positively and therefore do not need to be
identified. However, the multiplicity of protons at mid-rapidity is
small compared to that of pions and, therefore, including protons does
not significantly change the resolution.

\begin{figure}[t]
\centering
\mbox{
\mfigure[
{\it RQMD v2.4 prediction for elliptic flow using $\pi^+, \pi^-$ and
protons within 2.5~$\leq |y| \leq$~4.0.}  ] {
\epsfig{figure=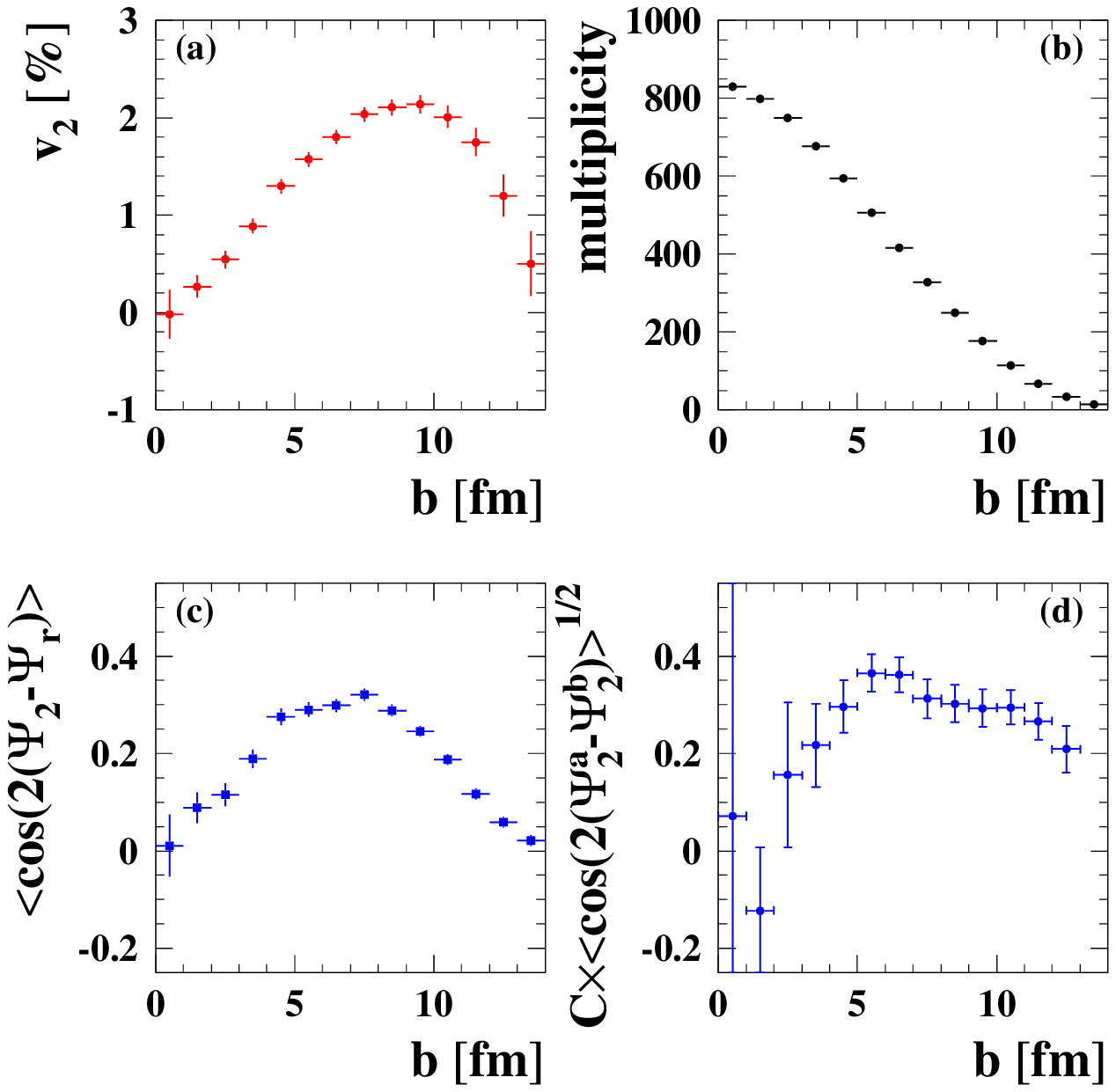,width=.48\textwidth}
\label{resv2ftpc}
}
\mfigure[
{\it RQMD v2.4 prediction for directed flow using $\pi^+, \pi^-$ and
protons within 2.5~$\leq |y| \leq$~4.0.}  ] {
\epsfig{figure=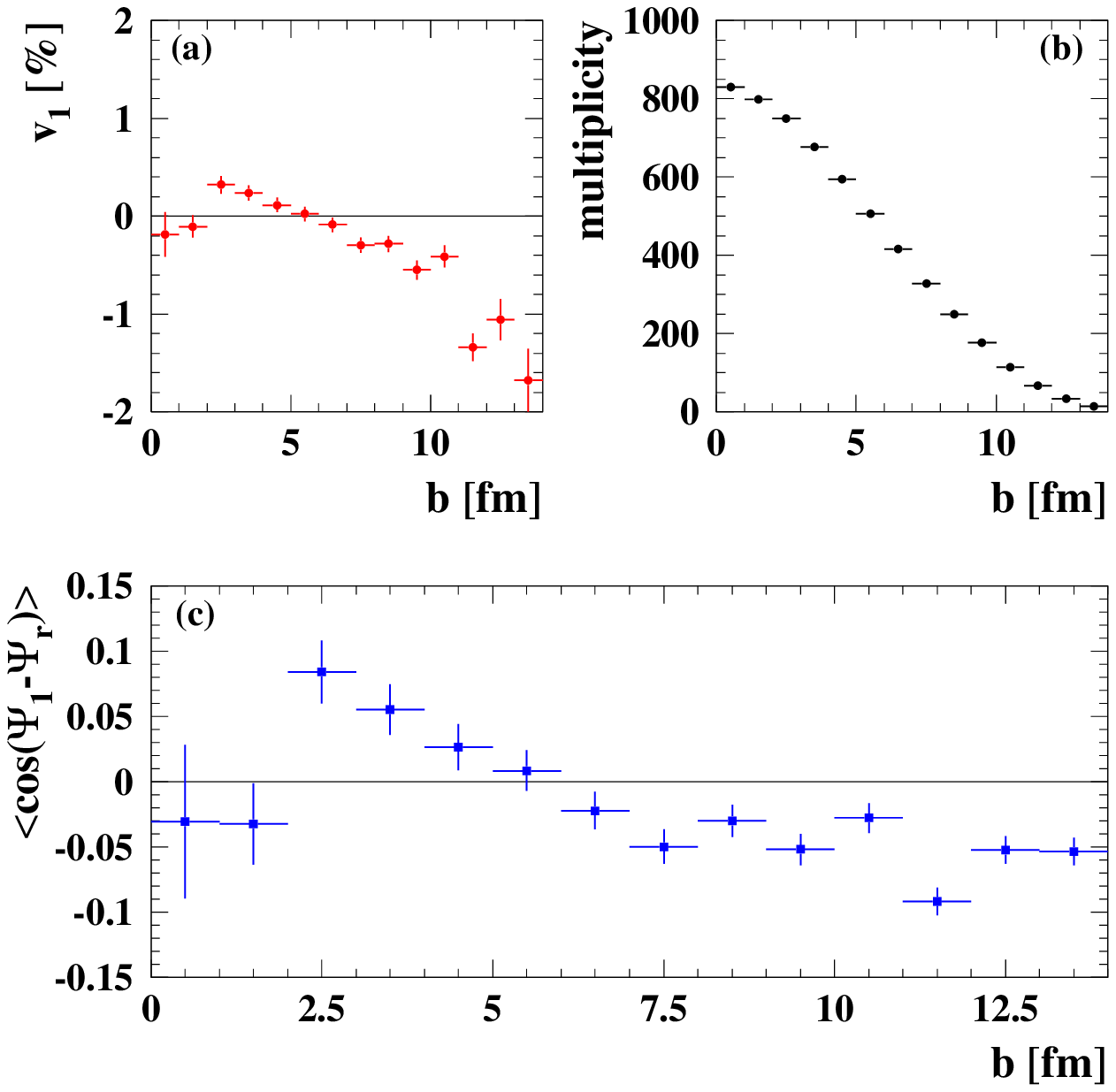,width=.48\textwidth}
\label{resv1ftpc}
}
}
\end{figure}
In Fig.~\ref{resv2ftpc}a, $v_2$ integrated over the FTPC rapidity
region is shown versus the impact parameter $b$.  For the FTPCs
the $\pi^+, \pi^-$ and protons are combined. It was shown in
Fig.~\ref{vxrx_rqmd}e that $v_2$ is relatively flat as a function of
rapidity and its magnitude is therefore comparable in the FTPC and TPC
regions.  Fig.~\ref{resv2ftpc}b shows the corresponding multiplicity
as a function of $b$ for the combined FTPCs.  These quantities lead to
a resolution for $v_2$, calculated using the true reaction plane,
as shown in Fig.~\ref{resv2ftpc}c.  The resolution for $v_2$ which can
be obtained in the STAR FTPCs using subevents is shown in
Fig.~\ref{resv2ftpc}d. If only one FTPC would be used this resolution
would be approximately $\sqrt 2$ smaller.

Using $v_2$ the event plane can be determined, however the sign of
$v_2$ is not determined relative to ${\bf b}$.  This sign could be
determined from $v_2$ relative to $\Psi_1$. Fig.~\ref{vxrx_rqmd}c
shows that around mid rapidity $v_1$ is maximally 0.5\% which makes
$\Psi_1$ extremely hard to measure. From Fig.~\ref{vxrx_rqmd}a
and~\ref{vxrx_rqmd}c it is clear that the best region to measure $v_1$
is at forward rapidity.  Fig.~\ref{resv1ftpc}a shows $v_1$ integrated
over the FTPC rapidity region, versus $b$.  As for $v_2$, the $\pi^+,
\pi^-$ and protons are combined. This decreases the magnitude of $v_1$
because their signs are opposite but the FTPCs are not able to
separate these particles. At large $b$ the magnitude of $v_1$ becomes
$\approx$ 1\% and, although this is already hard to measure, also the
multiplicity decreases rapidly at large $b$. This leads to negligible
resolution for $v_1$ at all values of $b$, which is shown in
Fig.~\ref{resv1ftpc}c.

\section{Conclusion}
We have investigated the feasibility of reconstructing the event
plane.  Both Fig.~\ref{restpc} and Fig.~\ref{resv2ftpc} show that it
is possible to determine the second harmonic event plane and calculate
$v_2$ within STAR, assuming the RQMD predictions (multiplicity
distribution, magnitude of $v_2$) are correct. For $v_2$ both the TPC
or the FTPCs can be used. This would initially provide a cross check
and later combining both detectors would increase the resolution. For
this study we only need the momenta of the charged hadrons and thus
anisotropic flow could be one of the first results from STAR. For
future analyses it would be good to have particle identification.
Because it is important to study the dependence of $v_2$ as a function
of $b$~\cite{sorge98} we would like to have 10 centrality bins,
which would be possible with $1\;000\;000$ minimum bias events.

\section{Acknowledgments}
We would like to thank the other members of the STAR LBNL Soft Hadron
Group and in particular the group leader Nu Xu for help with this
work.



\end{document}